# Photometry and astrometry of SS Leo Minoris during the 2006 October superoutburst

Jeremy Shears, David Boyd, Tom Krajci, Robert Koff, John R. Thorstensen & Gary Poyner

## Abstract

We report unfiltered CCD observations of the first confirmed superoutburst of the SU UMa-type dwarf nova SS LMi in October 2006. From a quiescent magnitude of ~21.7 it rose to 16.2, an outburst amplitude of ~5.5 magnitudes. It declined at 0.17 mag d$^{-1}$ for 5 days before slowing to 0.11 mag d$^{-1}$ for a further 3 days. The light curve revealed common superhumps with a peak-to-peak amplitude of 0.3 magnitude, which decayed and then re-grew concurrently with the change in decline rate. These were followed by a phase-changing transition to late superhumps. Analysis of these observations has revealed evidence for an orbital period of 0.05572(19) d and a common superhump period of 0.05664(2) d, giving a fractional superhump period excess of 0.017(5). From astrometry of SS LMi in outburst we have established for the first time its correct position as RA 10h 34m 05.85(1)s, Dec +31° 08' 00.00(18)" (J2000). The position commonly given for SS LMi is that of a nearby star.

## Introduction

Dwarf novae are a class of cataclysmic variable stars in which a white dwarf primary accretes material from an orbiting secondary star via Roche lobe overflow. The secondary is usually a late-type main-sequence star. In the absence of a significant white dwarf magnetic field, material from the secondary is processed through an accretion disc before settling on the surface of the white dwarf. From time-to-time, as material builds up in the disc, thermal instability drives the disc into a hotter, brighter state causing an outburst in which the star brightens by several magnitudes [1]. Dwarf novae of the SU UMa family (UGSU) occasionally exhibit superoutbursts which last several times longer than normal outbursts and may be up to a magnitude brighter. During a superoutburst the light curve of a UGSU star is characterized by superhumps. These are modulations in the light curve which are a few percent longer than the orbital period and are thought to be caused by precession of an eccentric accretion disc [1].

## History of SS LMi

SS LMi was discovered in 1980 April by L. Zacs of the Radioastrophysical Observatory in Riga, Latvia [2]. The star was detected at a maximum brightness of 15.97V on April 12 and declined at an average of 0.16 mag d$^{-1}$ over a period of 7 days. After 1980 May 5 the star was not visible, so it was fainter than B~18.5 and V~17.5. Zacs and his co-worker, A. Alksnis, examined over 270 plates taken at Riga between March 1970 and May 1979, but could not detect any other outbursts. They commented that there appeared to be no trace of the star on the Palomar Observatory Sky Survey (POSS) plates, implying a faint quiescent magnitude and thus a large outburst amplitude. On this basis they speculated that the star was either an extragalactic nova or an unusual dwarf nova. The General Catalogue of Variable Stars (GCVS) lists SS LMi as a possible type A (fast) nova presumably based on Alksnis & Zacs comment [3].

Several reports of SS LMi being detected at magnitude 17 to 18 exist in the literature [4, 5, 6]. However, these reports indicate that there is uncertainty as to whether the correct star was identified. As we shall discuss later, we have now shown that the position of SS LMi quoted in most references is incorrect and it is therefore likely that these reports relate to a nearby field star.

We searched the AAVSO International Database for evidence of other outbursts without success. The Database contains more than 600 observations between 1994 April and 2006 October, but all are negative, i.e. "fainter than" observations. Furthermore, the vast majority of these are visual estimates having a limiting magnitude between 14 and 15. Hence, even if there had been further outbursts during this period with the same maximum brightness as the 1980 outburst, it is likely that visual observers would have missed them.

SS LMi was added to the British Astronomical Association Variable Star Section (BAA-VSS) Recurrent Objects Programme in 1994 [7]. This programme was set up as a joint project between the BAA-VSS and *The Astronomer* magazine specifically to monitor poorly studied eruptive stars of various types where outbursts occur at intervals of greater than 1 year.

**Detection and course of the outburst**

The outburst reported here was first detected by JS on 2006 Oct 24.177 at 16.2C (Figure 1) and is the first confirmed outburst for more than 26 years [8]. Over the next 8 days, 9 time-series runs were conducted yielding 792 individual CCD images and totalling 20.8 hours of photometry. Table 1 summarises the instrumentation used and Table 2 contains a log of the time-series runs. Unfortunately each run was necessarily rather short due to the poor position of this object during late October in the pre-dawn sky. Raw images were dark-subtracted and flat-fielded before being measured using aperture photometry relative to one or more comparison stars with photometry given by Henden [12]. JS used GSC2511-0528 (13.486V) as the comparison star, TK employed a comparison star ensemble of GSC2511-0222 and GSC2511-0298 (15.163V, 14.211V) and RK used an ensemble of the same two comparison stars along with GSC2511-0929 (15.242V). In the case of JS typical photometric errors were 0.09 mag, whereas for TK and RK they were 0.05 mag.

| Observer | Telescope diameter (m) | Telescope type | CCD camera |
|---|---|---|---|
| JS | 0.10 | Fluorite refractor | Starlight Xpress SXV-M7 |
| TK | 0.28 | SCT | SBIG ST-7E |
| RK | 0.25 | SCT | Apogee AP-47 |
| DB | 0.36 | SCT | Starlight Xpress  SXV-H9 |

**Table 1: Instrumentation used**

| Run number | Date in 2006 (UT) | Start time (JD) 2454000+ | Duration (h) | No. of images | Mean mag (C) | Observer |
|---|---|---|---|---|---|---|
| 1 | Oct 24 | 32.677 | 1.5 | 80 | 16.27 | JS |
| 2 | Oct 26 | 34.931 | 2.2 | 106 | 16.68 | TK |
| 3 | Oct 27 | 35.926 | 2.4 | 128 | 16.88 | TK |
| 4 | Oct 27 | 35.937 | 1.9 | 51 | 16.89 | RK |
| 5 | Oct 28 | 36.918 | 2.3 | 125 | 17.05 | TK |
| 6 | Oct 28 | 36.922 | 2.3 | 62 | 17.01 | RK |
| 7 | Oct 29 | 37.912 | 2.5 | 93 | 17.18 | TK |
| 8 | Oct 31 | 39.909 | 2.8 | 79 | 17.41 | TK |
| 9 | Nov 1 | 40.903 | 2.9 | 68 | 17.52 | TK |

**Table 2: Log of time-series observations**

Figure 2 shows the light curve of the outburst. From maximum light at 16.2C on Oct 24, the outburst declined at an average rate of 0.17 mag d$^{-1}$ for the first 5 days. This is typical of a dwarf nova in decline and is similar to the decline of SS LMi during the 1980 April outburst. Over the next 3 days the rate of decline slowed to 0.11 mag d$^{-1}$.

**Position and identity of SS LMi**

Using Astrometrica [9] and the USNO CCD Astrograph Catalog release 2, we obtained the following mean position for SS LMi during the outburst:
RA 10h 34m 05.85s +/- 0.01s, Dec +31° 08' 00.00" +/- 0.18" (J2000)

This is ~10" south-east of the position listed in many references including the GCVS and Catalogue and Atlas of Cataclysmic Variable Stars [10]. To try to understand the reason for this discrepancy, we examined the POSS2 Blue plate in the Digitized Sky Survey [11], a section of which ~2.3' square is shown in Figure 3. We have labelled as SS LMi the object at the above position which we observed in outburst. C1 and C2 are two stars which appeared to be constant during our observations. Photometry by Henden [12] gives the V-band magnitude of C1 as 18.53. The position commonly listed for SS LMi corresponds to the position of C2 and we therefore conclude that this star may have been incorrectly identified as SS LMi in previous publications. In their study of suspected old novae, Robertson et al. [13] report B-band photometry, but include an inverted finding chart marking star C1 as the variable. Sproats et al. [14], who report J and K-band measurements of SS LMi, may also have used an incorrect position. Given the confusion about correctly identifying SS LMi, we suggest that previously reported observations of SS LMi in quiescence are treated with caution.

**Detection of superhumps**

The time-series photometry data obtained immediately after discovery of the outburst on Oct 24 (Figure 4a) shows a hump-like feature with an amplitude of ~0.3 mag in the light curve. However, the run was curtailed due to the onset of dawn and it was not long enough to definitively characterise this as a superhump. Notification of this discovery was quickly disseminated by email on the *baavss-alert* and *cvnet-outburst* distribution lists.

Observations on Oct 26 were inconclusive as there was considerable scatter and no obvious humps (Figure 4b). However, the presence of superhumps was confirmed on Oct 27 (Figure 4c), now with a peak-to-peak amplitude of ~0.25 mag, revealing for the first time the star's UGSU identity.

Superhumps were again detected on Oct 28 and 29 (Figure 4d and 4e), but their amplitude was smaller (~0.2 mag and ~0.15 mag respectively) and less well defined due to the presence of considerable flickering. However by Oct 31 the superhumps had grown in amplitude again (~0.25 mag) and by Nov 1 had regained their original amplitude of ~0.3 mag (Figure 4f and 4g).

The reduction in the rate of decline noted above is often associated with such a re-growth in superhumps which takes place at the same time. This has been observed in many UGSU stars, especially those showing very long intervals between outbursts (i.e. of the order of hundreds to thousands of days)[15].

**Superhump and orbital periods**

In order to study the superhump behaviour, we first extracted the times of each resolvable superhump maximum from the individual light curves by fitting a quadratic function to the light curve around the time of maximum. Although there was considerable scatter in the data points due to flickering and the faintness of the variable, this method gave reasonable results. Times of 12 superhump maxima were found. These were then used to assign superhump cycle numbers which best fitted the assumption of a constant superhump period. We found that the 10 maxima in the period Oct 24 to Oct 31 fitted well a constant superhump period $P_{sh}$ = 0.05664(2) d with a superhump maximum ephemeris HJD 2454032.726(2) + 0.05664(2) * E. The dates and measured times of superhump maximum, the assigned superhump cycle numbers and the O-C (Observed-Calculated) residuals relative to this superhump maximum ephemeris are listed in Table 3 while the O-C residuals are plotted in Figure 5. This shows that the 2 superhumps on Nov 1 had experienced a large change in phase of 0.56 (or -0.44).

| Date | Time of maximum (HJD) 2450000+ | Superhump cycle no | O-C (cycles) |
|--------|-------------------------------|--------------------|------------------|
| Oct 24 | 4032.7270+/-0.0027 | 0 | 0.011+/-0.047 |
| Oct 27 | 4035.9539+/-0.0018 | 57 | -0.014+/-0.031 |
| Oct 27 | 4035.9547+/-0.0019 | 57 | -0.001+/-0.034 |
| Oct 28 | 4036.0102+/-0.0010 | 58 | -0.021+/-0.017 |
| Oct 28 | 4036.9747+/-0.0020 | 75 | 0.009+/-0.035 |
| Oct 28 | 4036.9748+/-0.0032 | 75 | 0.010+/-0.057 |
| Oct 29 | 4037.9375+/-0.0020 | 92 | 0.008+/-0.035 |
| Oct 29 | 4037.9934+/-0.0029 | 93 | -0.004+/-0.052 |
| Oct 31 | 4039.9192+/-0.0022 | 127 | -0.002+/-0.038 |
| Oct 31 | 4039.9761+/-0.0020 | 128 | 0.003+/-0.035 |
| Nov 1 | 4040.9152+/-0.0019 | 144 | 0.583+/-0.033 |
| Nov 1 | 4040.9697+/-0.0011 | 145 | 0.546+/-0.020 |

**Table 3: Timing of superhump maxima**

Following accepted practice, we interpret this behaviour in terms of a common superhump regime [1, 16] operating up to cycle ~130 with period $P_{sh}$ = 0.05664(2) d followed by a large phase change marking transition to late superhumps. Such anti-phased late superhumps have been found in a variety of UGSU stars including IY UMa, V1159 Ori, ER UMa, TT Boo and WZ Sge [1, 17, 18, 19, 20]. At present, the cause of late superhumps is not fully understood, although there is a growing body of evidence that it is associated with the hot spot formed where the material streaming from the secondary impacts the edge of the accretion disc [17, 19]. It has been proposed that more energy is released upon impact when the stream falls furthest; this would occur when the smaller side of the accretion disc is nearest the secondary, which is the opposite to the situation for common superhumps and so would explain the ~0.6 cycle phase change [1]. We also note that this transition to late superhumps occurs at the same time as the relative brightening in the light curve and the re-growth of superhumps described earlier.

As all datasets were longer than the superhump period, a linear fit was subtracted from each dataset before carrying out a period analysis of all the data from Oct 24 to Oct 31 using the ANOVA algorithm in Peranso [21] (data from Nov 1 were excluded due to the phase shift). This gave the power spectrum in Figure 6a where we interpret the peak at 17.64(6) cd$^{-1}$ as being due to common superhumps. The error estimate is derived using the Schwarzenberg-Czerny method [22]. While the corresponding period of 0.05668(19) d is consistent with our earlier finding, it is not unusual for O-C analysis to give a more accurate method of tracking periodic waves in dwarf novae as it is less troubled by changes in amplitude than period analysis techniques [23]. In view of the consistency of the O-C analysis, in this case we take our value of $P_{sh}$ from that analysis. The associated 1 cd$^{-1}$ aliases of the superhump signal are expected since most of the data were obtained at a similar longitude. The power spectrum also shows signals at 35.28 cd$^{-1}$, the first harmonic of the superhump frequency, plus 1 cd$^{-1}$ aliases. A phase diagram of the data folded on $P_{sh}$ is shown in Figure 7. The average peak-to-peak amplitude of the superhump signal was 0.18 mag. Besides the main peak this shows a secondary peak at phase ~0.6 which may be related to the growth of late superhumps since, as we noted above, these are displaced in phase relative to common superhumps by a similar amount.

We removed the superhump signal by pre-whitening the data with 17.64 cd$^{-1}$ and repeated the period analysis. This produced the power spectrum shown in Figures 6b and 6c. We interpret the peak at 17.95(6) cd$^{-1}$ as the orbital signal with $P_{orb}$ = 0.05572(19) d, although the signal is weak. Its peak-to-peak amplitude was 0.06 mag. At 80.2 min, this is one of the shortest orbital periods for a UGSU star on record [24]. The fractional superhump period excess $\varepsilon$ = ($P_{sh}$ - $P_{orb}$) / $P_{orb}$ is 0.017(5), consistent with values of $\varepsilon$ for other UGSU-type dwarf novae with short orbital periods listed for example in [25]. Using the relationship $\varepsilon$ = 0.18q + 0.29q$^2$ from [26], we find the mass ratio q = 0.083(22).

Period analysis of the Nov 1 data gave a poorly-determined late superhump frequency of 18.0(7) cd$^{-1}$ (period 0.0555(22) d). The phase diagram of the Nov 1 data is shown in Figure 8.

As a check, the data was reanalysed using the Data Compensated Discrete Fourier Transform (DCDFT) method and this produced results consistent with those above within the estimated uncertainties.

A sub-class of UGSU stars is recognised by some researchers which are more highly evolved systems, having short orbital periods, long intervals between outbursts and exceptionally large outburst amplitudes [23, 27]. This sub-class is usually referred to as "UGWZ" after its prototype WZ Sge. In view of its short orbital period, we have considered whether SS LMi might belong to the UGWZ class of dwarf novae, but have decided this is unlikely as it lacks several of the distinguishing characteristics of this class. The initial decline rate was not as large and double-peaked outburst orbital humps, sometimes referred to as early superhumps, were not observed (although it is possible that we missed this early stage of the outburst). Several UGWZ stars exhibit re-brightenings, or "echo outbursts", after the main outburst, however we cannot comment on possible re-brightenings of SS LMi as we were unable to observe it during this period. Finally, we cannot draw any specific conclusions about the outburst interval because of the lack of earlier positive observations.

By contrast, we note the similarity of SS LMi's orbital period and superhump period excess with V1108 Her (also called "Var Her 04") [28]. This is an enigmatic dwarf nova with some characteristics in common with the hydrogen-burning "period bouncers", dwarf novae whose orbital periods are lengthening after evolving through the period minimum.

## SS LMi at quiescence

We followed SS LMi towards quiescence. On 2006 Nov 17.07, 24 days after the outburst was first detected, a 19 min unfiltered stacked exposure by DB recorded the object at magnitude ~21.2C, while on Dec 17.03 it was just measurable in a 32 min unfiltered stacked exposure at magnitude ~21.7C. A CCD V-band image (Figure 9) taken by JRT on 2000 Apr 4 using a 2.4 m telescope revealed nothing at the position of SS LMi. Stars as faint at 21.0 can easily be seen in this image and from this we conclude that SS LMi reaches V > 21.5 in quiescence. This means that the outburst amplitude of the current outburst was probably at least 5.3 magnitudes. However, it is possible that the star was already in decline when we first detected it at 16.2C; taking the maximum brightness of the 1980 outburst implies an outburst amplitude of at least 5.5 magnitudes.

## Conclusion

We have reported unfiltered CCD observations of the first confirmed superoutburst of the UGSU-type dwarf novae SS LMi in October 2006. From a quiescent magnitude of ~21.7 it rose to 16.2, an outburst amplitude of ~5.5 magnitudes. It declined at 0.17 mag d$^{-1}$ for 5 days before slowing to 0.11 mag d$^{-1}$ for a further 3 days. The light curve displayed 0.3 magnitude peak-to-peak superhumps which decayed and then re-grew concurrently with the change in decline rate. These were followed by a phase-changing transition to late superhumps. Analysis of these observations has revealed evidence for an orbital period of 0.05572(19) d, one of the shortest on record for a UGSU star, and a common superhump period of 0.05664(2) d, giving a fractional superhump period excess of 0.017(5). From astrometry of SS LMi in outburst we have established for the first time its correct position as RA 10h 34m 05.85(1)s, Dec +31° 08' 00.00(18)" (J2000). The position commonly given for SS LMi is that of a nearby star.

## Acknowledgements

The authors gratefully acknowledge the use of observations from the AAVSO International Database contributed by observers worldwide, the use of the Digitized Sky Surveys produced at the Space Telescope Science Institute under U.S. Government grant NAG W-2166, and the use of SIMBAD, operated through the Centre de Données Astronomiques (Strasbourg, France). We are indebted to Drs Boris Gaensicke (University of Warwick, UK) and Chris Lloyd (Open University, UK) for helpful advice during the preparation of this paper and to the referees for their constructive comments which have improved the paper.

**Addresses:**
JS: "Pemberton", School Lane, Bunbury, Tarporley, Cheshire, CW6 9NR, UK
[bunburyobservatory@hotmail.com]
DB: 5 Silver Lane, West Challow, Wantage, Oxon, OX12 9TX, UK
[drsboyd@dsl.pipex.com]
TK: CBA New Mexico, PO Box 1351 Cloudcroft, New Mexico 88317, USA
[tom_krajci@tularosa.net]
RK: 980 Antelope Drive West, Bennett, CO 80102, USA
[bob@AntelopeHillsObservatory.org]
JRT: Dept of Physics & Astronomy, Dartmouth College, 6127 Wilder Laboratory, Hanover,
NH 03755, USA [thorsten@partita.dartmouth.edu]
GP: 67 Ellerton Road, Kingstanding, Birmingham, B44 0QE, UK
[garypoyner@blueyonder.co.uk]

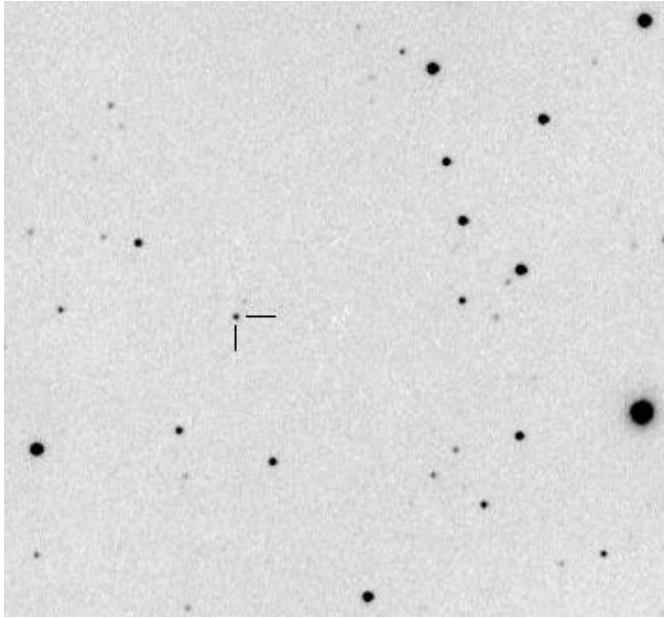

Figure 1: SS LMi in outburst at 16.2C on 2006 Oct 24.177
11' x 11' with N at top, E to left (J Shears)

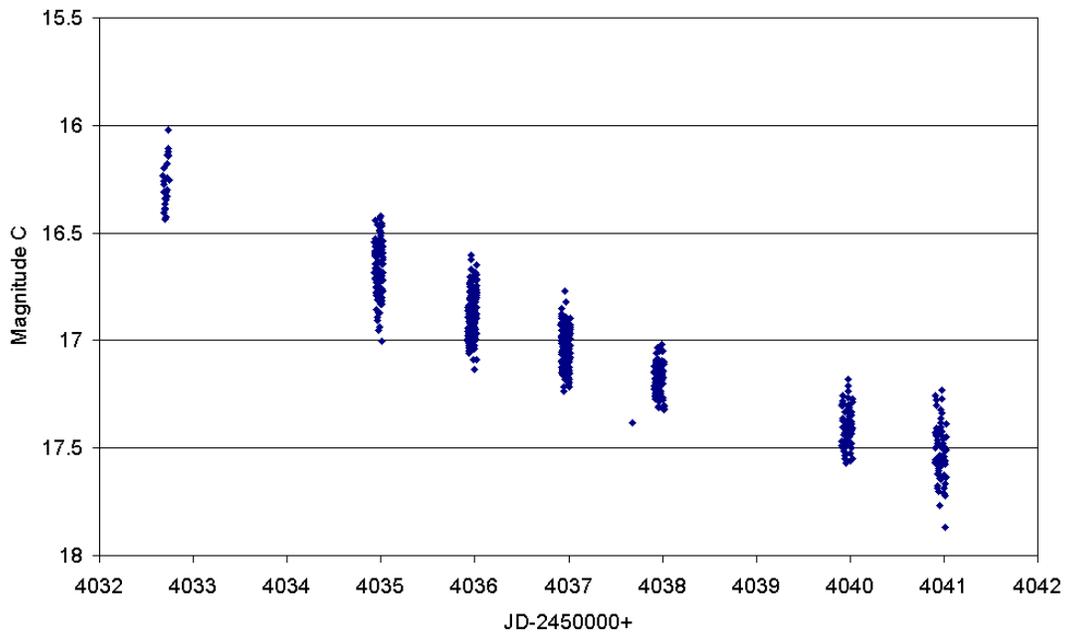

Figure 2: Unfiltered light curve of the 2006 October outburst

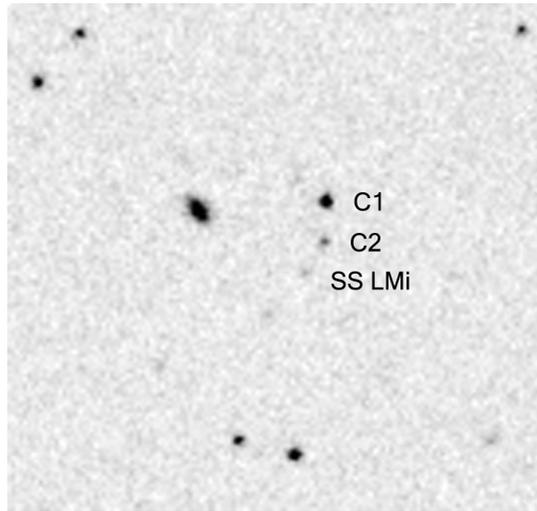

Figure 3: Section of POSS2 Blue plate showing the correct identification of SS LMi
~2.3 arcmin square with N at top, E to left

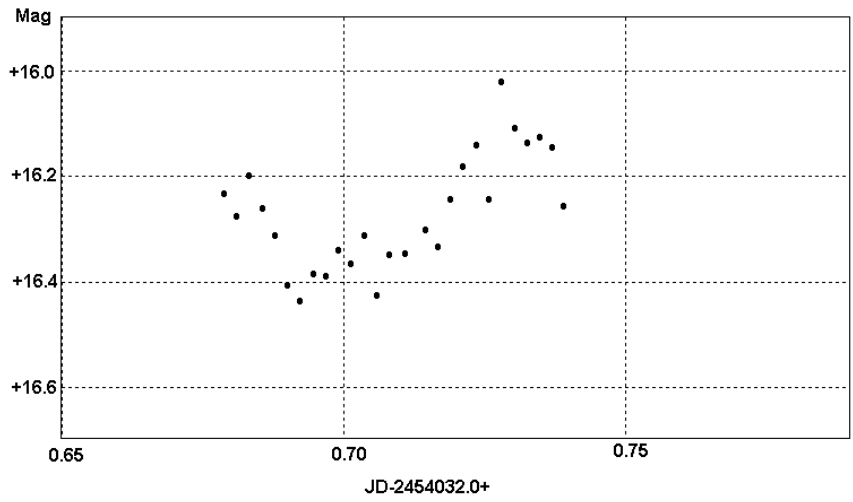

**(a) Oct 24**

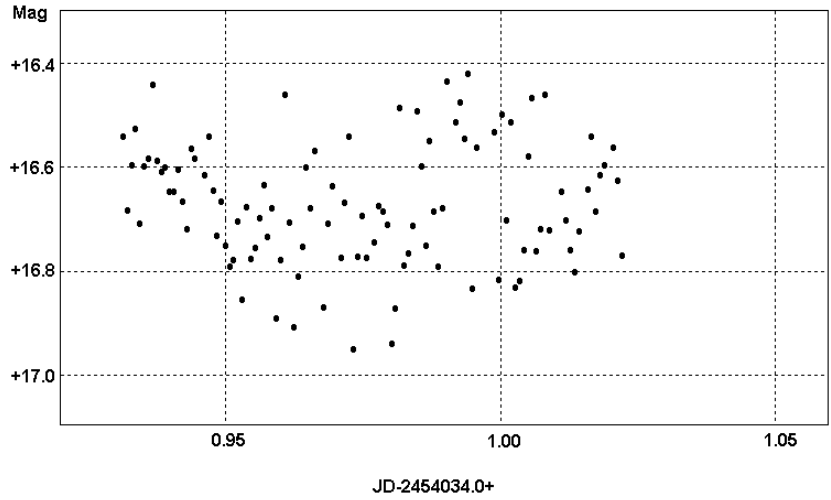

**(b) Oct 26**

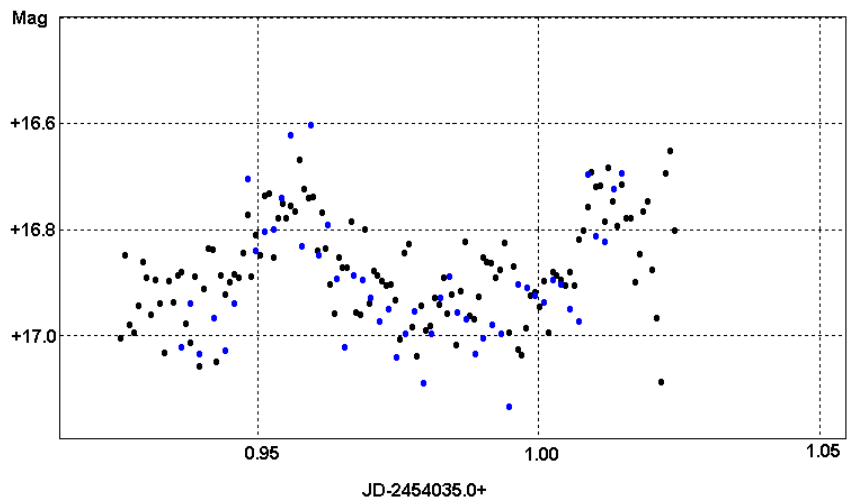

**(c) Oct 27**

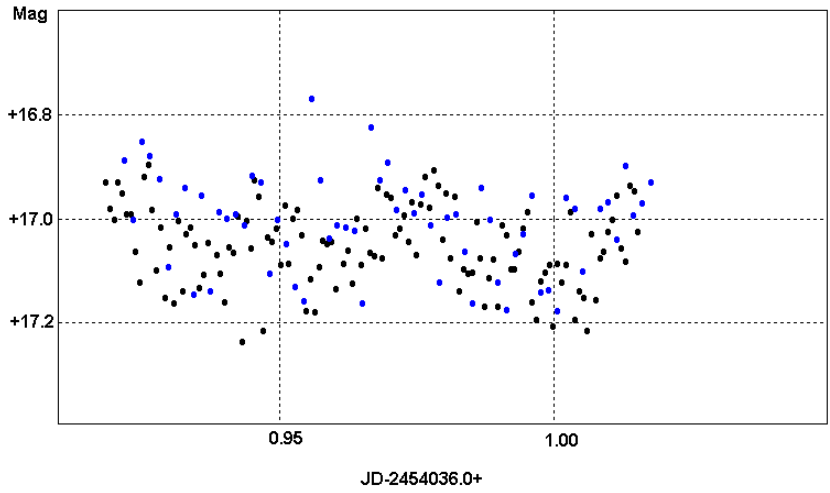

**(d) Oct 28**

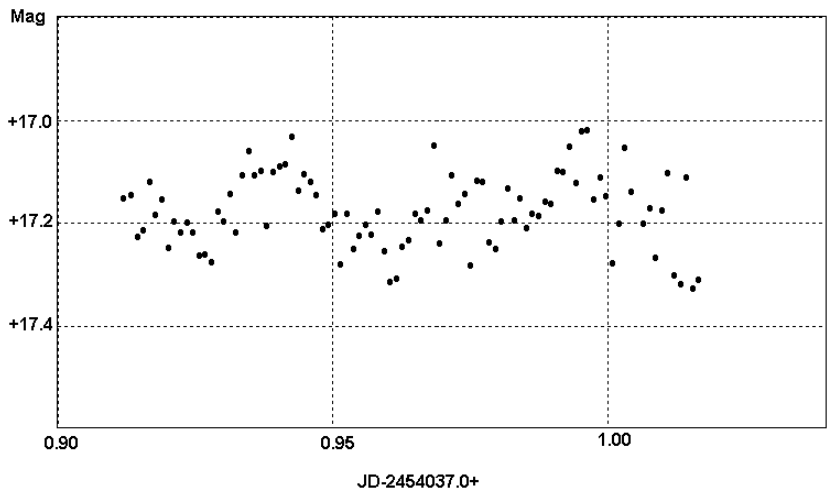

**(e) Oct 29**

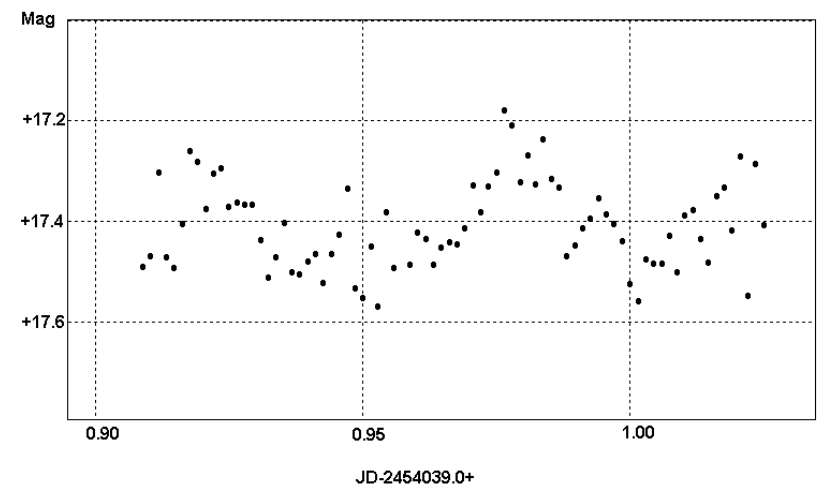

**(f) Oct 31**

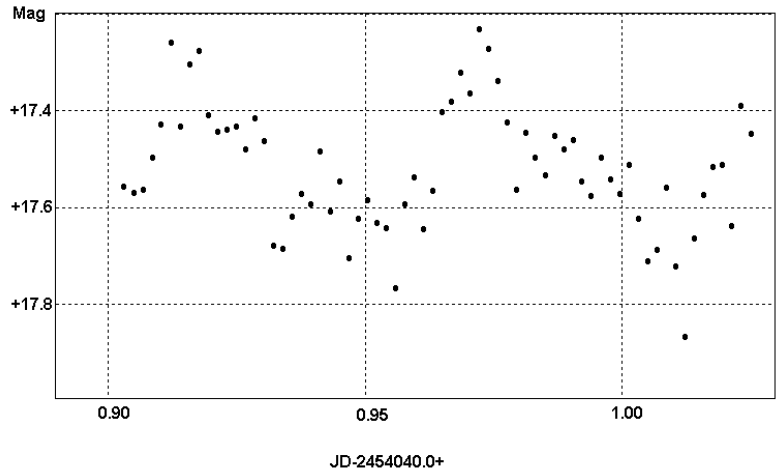

**(g) Nov 1**

Figure 4: Time series data
(J Shears, T Krajci and R Koff)

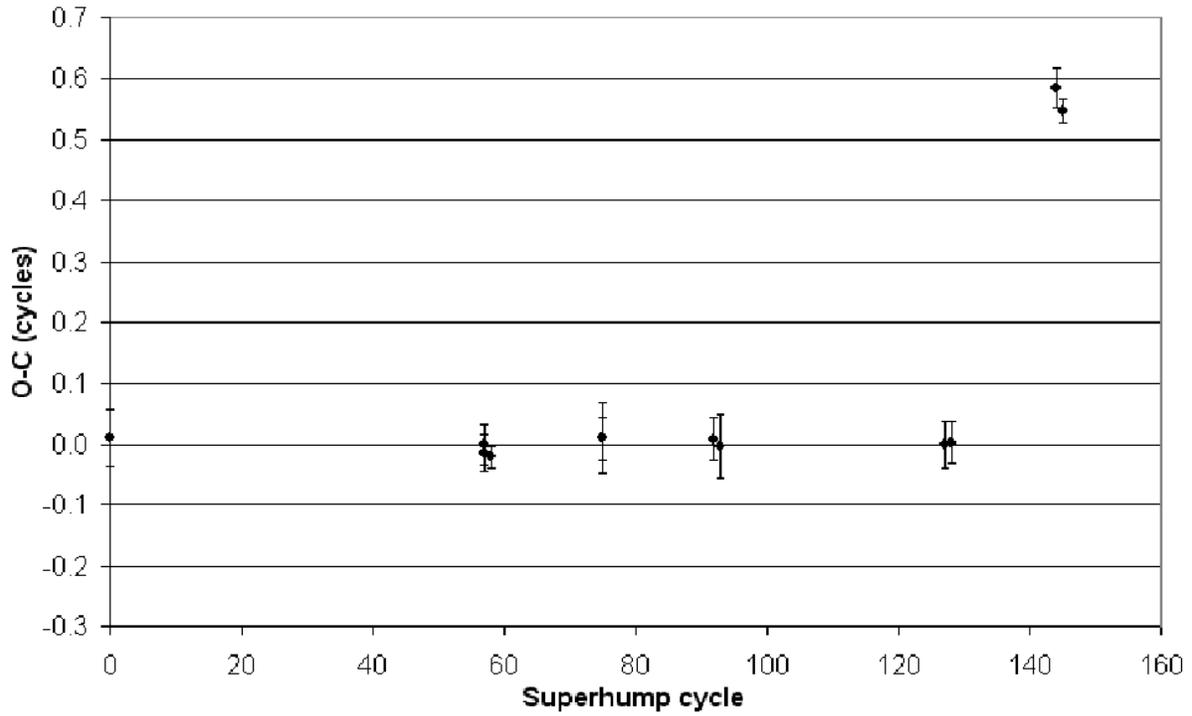

Figure 5: O-C residuals of times of superhump maximum relative to the ephemeris in this
paper

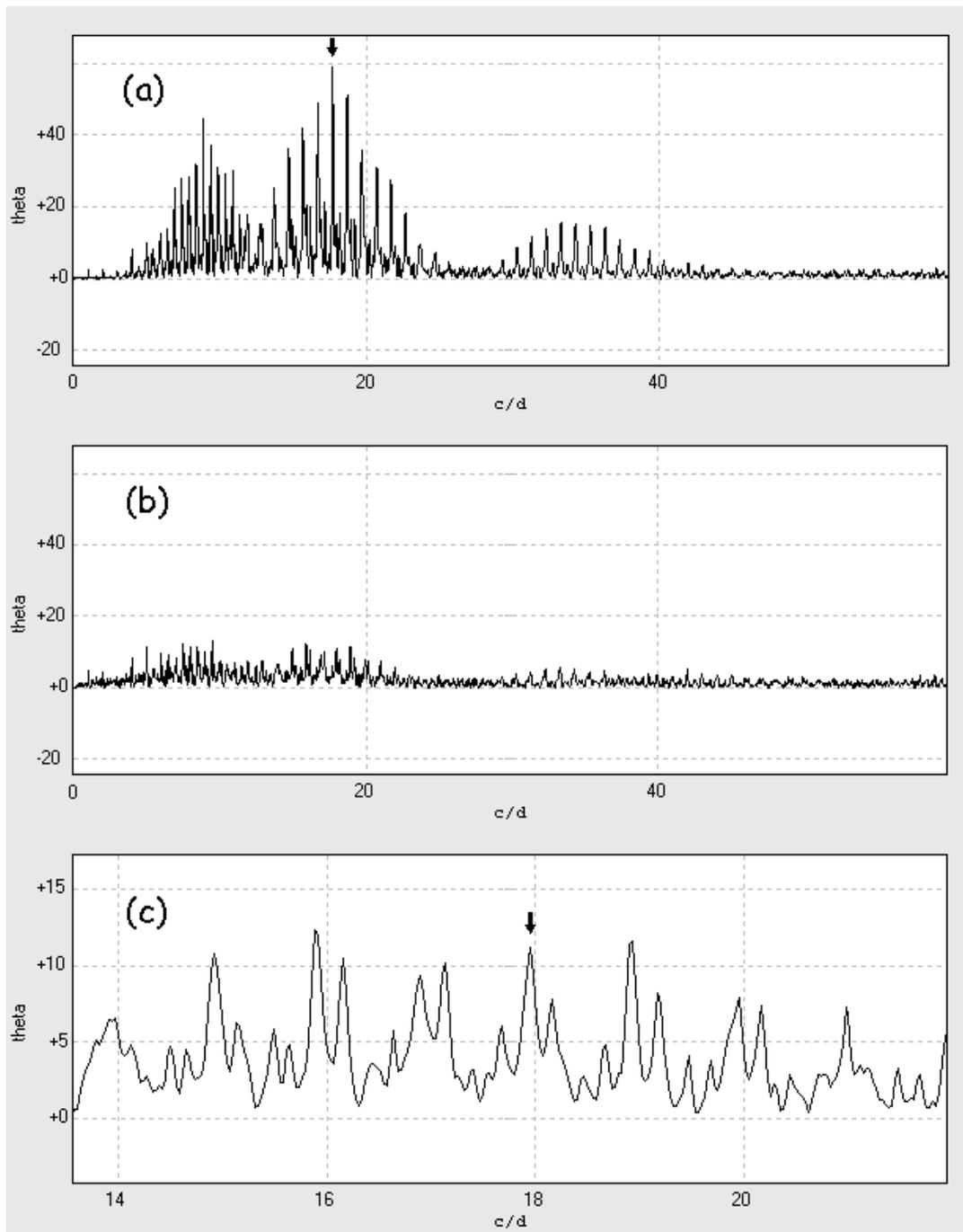

Figure 6: (a) Power spectrum of all data from Oct 24 to Oct 31 indicating common
superhump signal at 17.64 cd$^{-1}$
(b) Power spectrum after pre-whitening with 17.64 cd$^{-1}$
(c) Enlarged detail from (b) indicating orbital signal at 17.95 cd$^{-1}$

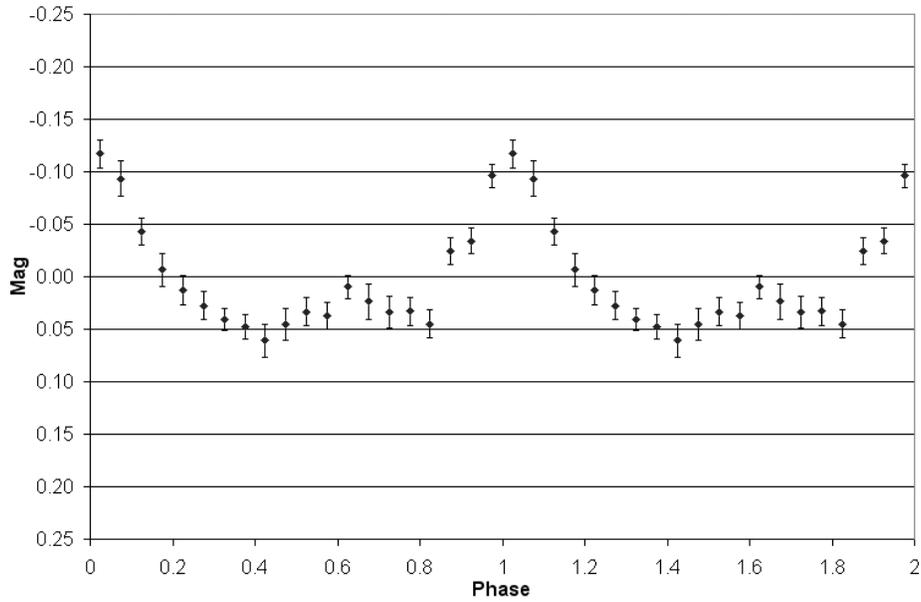

Figure 7: Phase diagram of all data from Oct 24 to Oct 31 folded on $P_{sh}$ = 0.05664 d with standard errors on each bin, showing 2 cycles

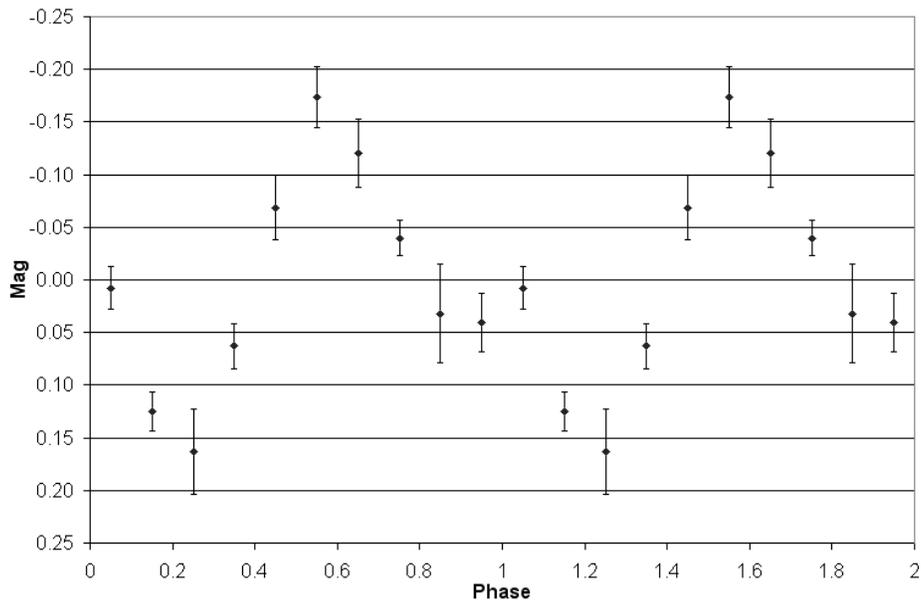

Figure 8: Phase diagram of data on Nov 1 folded on $P_{sh}$ = 0.0555 d with standard errors on each bin, showing 2 cycles

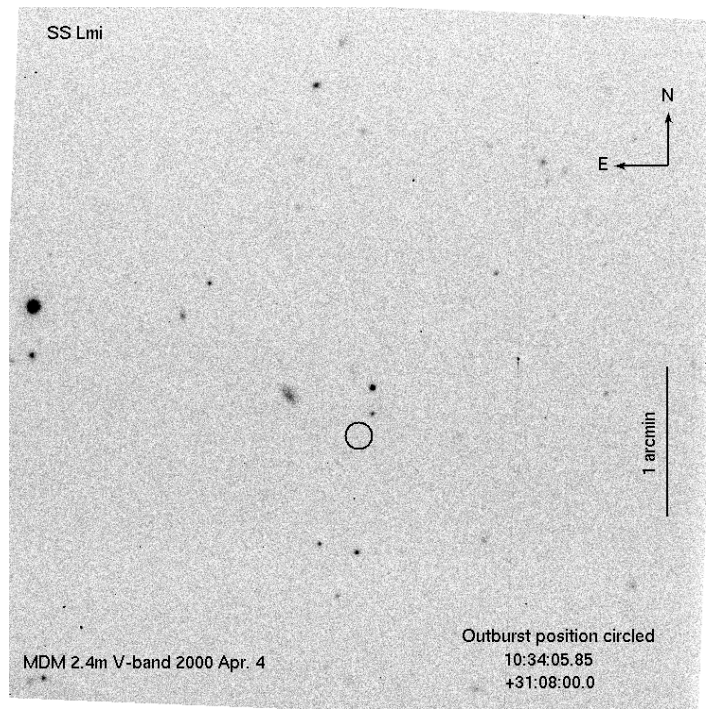

SS Lmi

MDM 2.4m V-band 2000 Apr. 4

Outburst position circled
10:34:05.85
+31:08:00.0

1 arcmin

Figure 9: V-band image of SS LMi location taken on 2000 Apr 4 with a 2.4m telescope,
field ~4.7 arcmin wide, with N at top, E to left
(J Thorstensen)